\documentclass[aps,prl,twocolumn,superscriptaddress,showpacs,preprintnumbers,bibliography]{revtex4-1}
\usepackage[colorlinks,bookmarks=false,citecolor=blue,linkcolor=red,urlcolor=blue]{hyperref}
\usepackage{physics}
\input pdfcolor.tex
\usepackage{colortbl,amsthm,txfonts}
\usepackage{verbatim}
\usepackage{graphicx}
\usepackage{epsfig}
\usepackage{dcolumn}
\usepackage{bm}

\usepackage{amsmath,amssymb}

\makeatletter
\renewcommand*\env@matrix[1][*\c@MaxMatrixCols c]{%

\hskip -\arraycolsep
\let\@ifnextchar\new@ifnextchar
\array{#1}}
\makeatother
\usepackage{appendix}

\begin{document}

\title{Kagome Hubbard model away from the strong coupling limit: Flat band localization and non Fermi liquid signatures}
\author{Shashikant Singh Kunwar}
\affiliation{Department of Physics and Astronomy, University of Iowa, \\
Iowa City, Iowa 52242, United States of America}
\author{Madhuparna Karmakar}
\email{madhuparna.k@gmail.com}
\affiliation{Department of Physics and Nanotechnology, SRM Institute of Science and Technology,   \\ 
Kattankulathur, Chennai 603203, India}

\begin{abstract}
Taking cue from the recent experimental realization of metallic phases in Kagome materials we report the low 
temperature signatures and thermal scales of Kagome metals and insulators, determined in the framework 
of the Kagome Hubbard model, using a non perturbative numerical approach. In contrast to the existing consensus 
we establish a flat band localized insulator in the weak coupling regime which crosses over to a non Fermi liquid (NFL) 
metal at intermediate coupling, followed by a first order metal-Mott insulator transition in the strong coupling regime. We 
provide the first accurate estimates of the thermal scales of this model and analyze the NFL phases in terms of resilient 
quasiparticles and short range magnetic correlations. With our unprecedented access to the low temperature phases and 
sufficiently large system sizes, we  provide the essential benchmarks for the prospective experiments on the Kagome metal 
and insulators in terms of their thermodynamic, spectroscopic and transport signatures. 
\end{abstract}  

\date{\today}
\maketitle

\textit{Introduction:}
Electronic correlation and magnetic frustration are the two ingredients which promises 
a rich phase diagram of the candidate material \cite{dagotto_science2005,balents_natphys2010,haule_rmp2011,kawakami_prb2002,kotlier_prl2004,trembley_prl2006,kawakami_prl2008,georges_prx2021}. A paradigmatic model to bring forth the interplay between strong electronic correlation and magnetic frustration is the Kagome Hubbard model (KHM) \cite{maekawa_prl2005,tsunetsugu_prl2006,ohta_prb2011,kawakami_prb2013,zhu_prbl2021,janson_prb2021}. 
Extensively investigated in the context of spin-$1/2$ quantum magnets, \cite{ballou_prl2008,cheong_nature2002,udagawa_prl2017,shender_prl1992,rutenberg_prb1992,berlinsky_prb1993,zhitomirsky_prb2008,moessner_prl2013,pujol_prb2018,canals_prb2014} the true ground state in the strong coupling (Heisenberg limit) of the 
KHM continues to be debated till date, with the candidate low energy states being a Dirac spin liquid \cite{wen_prl2007,wen_prb2008,marston_prl2008,poilblanc_prb2013,pollmann_prx2017,xiang_prl2017,su_scibull2018}, 
a $Z_{2}$ spin liquid \cite{white_science2011,schollwock_prl2012,balents_natphys2012,hotta_natcom2013,schollwock_prb2015,wen_prb2017,moessner_prb2019}, a chiral spin liquid \cite{lhuillier_prl2012,weinstein_prb2013,sheng_scirep2014,ludwig_natcom2014,chen_prl2014,lauchili_prb2015,sheng_prb2015,sheng_prb2015_2,bernu_prl2017} and a valence bond solid \cite{zeng_jap1991,maleyev_prb2002,senthil_prb2003,huse_prb2007,auerbach_prl2004,vidal_prl2010,poilblanc_prb2010,schwandt_prb2010,misguich_prb2011}. At this regime of interaction the static magnetic structure factor hosts ``pinch points'' \cite{udagawa_prb2018,zhitomirsky_prb2008,shanon_prb2018,iqbal_prr2023}, a low but finite temperature bow-tie like feature salient to the magnetically frustrated materials such as, pyrochlore lattice \cite{gingras_science2001,bramwell_science2009,zaharko_prb2018}. 

Contrary to the extensive investigation carried out to understand the Heisenberg limit of the 
KHM, the investigation of the weaker correlation regimes are far and few.  The discovery of Kagome metals 
such as, Mn$_{3}$Sn \cite{higo_nature2015,parkin_sciadv2016,shin_natmat2017,otani_nature2019,behnia_natcom2019,hess_prb2019}, Fe$_{3}$Sn$_{2}$ \cite{wills_jpcm2009,wills_jpcm2011,zhang_advmat2017,checkelsky_nature2018,wang_nature2018,zhang_prl2018,wang_apl2019,tranquada_prl2019,kondo_prb2020}, Co$_{3}$Sn$_{2}$S$_{2}$ \cite{chen_natphys2018,lei_natcom2018,hasan_natphys2019,chen_science2019,shen_apl2019,analytis_natcom2020,madhavan_natcom2021}, Gd$_{3}$Ru$_{4}$Al$_{12}$ \cite{ochiai_prb2018,ochiai_jspj2019}, FeSn \cite{checkelsky_apl2019,ghimire_natmat2019,zhang_prb2020,mcguire_prm2019} and 
more recently AV$_{3}$Sb$_{5}$ series (A= K, Rb, Cs) \cite{wilson_prm2021,wilson_prl2020} however, have brought the KHM 
back into the focus. 

A consensus on the ground state and finite temperature physics of KHM, that emerges based on the dynamical 
mean field theory and its cluster variants (DMFT and CDMFT) \cite{tsunetsugu_prl2006,tsunetsugu_jpcm2007,asano_prb2016,hatsugai_prl2019,ohashi_prl2006}, determinant 
quantum Monte Carlo (DQMC) \cite{maekawa_prl2005,janson_prb2021,paiva_prb2023}, D$\Gamma$A \cite{held_rmp2018,held_prb2021,toschi_prb2016}, variational cluster approximation (VCA) etc. \cite{asano_prb2016} suggests that  the ground state the half filled KHM 
hosts a metal deep in the weak coupling regime and undergoes a first order metal-insulator transition (MIT) at a critical interaction 
$U_{c}$ that ranges between $5t-11t$ depending upon the choice of the numerical approach and the correlation effects retained therein. 
A recent density matrix renormalization group (DMRG) calculation showed two critical points corresponding to a translation symmetry 
broken insulator at $U_{c1} \sim 5.4t$ and a quantum spin liquid at $U_{c2} \sim 7.9t$ \cite{zhu_prbl2021}. 

In this letter,  we use a static path approximated (SPA) Monte Carlo technique,  extensively benchmarked and 
implemented in the context of quantum phases and phase transitions \cite{evenson_jap1970,avishai_nature2007,mk_pra2016,mk_pra2018,nyayabanta_prb2016,nyayabanta_prb2022,nyayabanta_jpcm2017,mk_prr2022,anamitra_prl2017,mk_prb2022,dagotto_prl2005,dagotto_prb2005}, to map out the low temperature phases and thermal scales of the KHM, 
over a range of the Hubbard interaction $U$. In order to take into account the spatial fluctuations of the fermionic correlations 
in this multiband system, over a reasonably large system size, we develop and implement a highly parallelizable cluster diagonalization scheme \cite{dagotto_pre2015} termed as the parallel traveling cluster approach (PTCA),  which gives us the required edge over the existing numerical 
techniques in terms of the computation cost (see supplementary materials (SM) for the details). 

Our key inferences based on the thermodynamic, spectroscopic and transport signatures of the half filled KHM includes: 
at low temperatures, $(i)$ we establish a gapless flat band localized non Fermi liquid insulator (FI-NFL), 
in the weak coupling regime of $0 < U \le U_{c1} \sim 3.6t$, $(ii)$ the intermediate 
and strong coupling regimes host a NFL metal with $U_{CM} \sim 4.0t$ demarcating the onset of antiferromagnetic 
correlations and $U_{c2} \sim 4.4t$ quantifying the MIT to an antiferromagnetic Mott insulator, respectively, $(iii)$ based on the experimentally 
accessible spectroscopic, transport and thermodynamic signatures we provide the first exact estimate of the thermal scales and ascertain 
the NFL characteristics of the underlying quantum phases, describable in terms of the resilient quasiparticles which survives the breakdown 
of the Fermi liquid theory.   

\textit{Model, method and indicators:}
Our starting Hamiltonian is the two-dimensional repulsive Hubbard model on a Kagome lattice, 
which reads as \cite{janson_prb2021,paiva_prb2023}, 
\begin{eqnarray}
\hat H & = & -\sum_{\langle ij \rangle, \sigma} t_{ij}(\hat c_{i, \sigma}^{\dagger}\hat c_{j, \sigma} + h. c. ) -\mu\sum_{i, \sigma} \hat n_{i\sigma} + U \sum_{i}\hat n_{i, \uparrow}\hat n_{i,\downarrow} \nonumber \\ 
\end{eqnarray}
where, $t_{ij} = t=1$ is the hopping integral between the nearest neighbors on a Kagome lattice and sets 
the reference energy scale of the problem, $U > 0$ corresponds to the on-site Hubbard repulsion. We work at 
half filling and the chemical potential $\mu$ is adjusted to maintain the same. The model is made numerically 
tractable via Hubbard-Stratonovich (HS) decomposition \cite{hs1,hs2} of the interaction term,  introducing a 
vector ${\bf m}_{i}(\tau)$ and a scalar $\phi_{i}(\tau)$ (bosonic) auxiliary fields at each site, which couples to 
the spin and the charge channels, respectively. The problem is addressed via static path approximated (SPA) 
Monte Carlo technique wherein the model is envisaged as an effective spin-fermion model with the random, 
fluctuating, ``classical'' background of the auxiliary fields coupled to the free fermions. The $\phi_{i}$ field is 
treated at the saddle point level as $\phi_{i} \rightarrow \langle \phi_{i}\rangle = \langle n_{i}\rangle U/2$ (where, 
$\langle n_{i}\rangle$ is the number density of the fermions), while the complete spatial fluctuations of ${\bf m}_{i}$ 
are retained. The fermionic correlators based on which the phases are characterized, includes: $(i)$ spectral 
function ($A({\bf k}, \omega)$), $(ii)$ single particle density of states (DOS) ($N(\omega)$), $(iii)$ local single 
particle DOS at the Fermi level ($N_{i}(0)$), $(iv)$ static magnetic structure factor ($S({\bf q})$), 
$(v)$ nearest ($\langle {\bf m}_{i}.{\bf m}_{j}\rangle$) and next nearest ($\langle {\bf m}_{i}.{\bf m}_{k}\rangle$) neighbor 
magnetic correlations, $(vi)$ magnetic order parameter ($m^{2}$), $(vii)$ optical conductivity ($\sigma(\omega)$) 
and $(viii)$ dc-conductivity ($\sigma_{dc}$) (see SM for details). The results presented in this letter corresponds to 
a system size of $3\times L^{2}$, with $L=18$, and are verified to be robust against finite system size effects.  
\begin{figure}
\begin{center}
\includegraphics[height=3.3cm,width=6.5cm,angle=0]{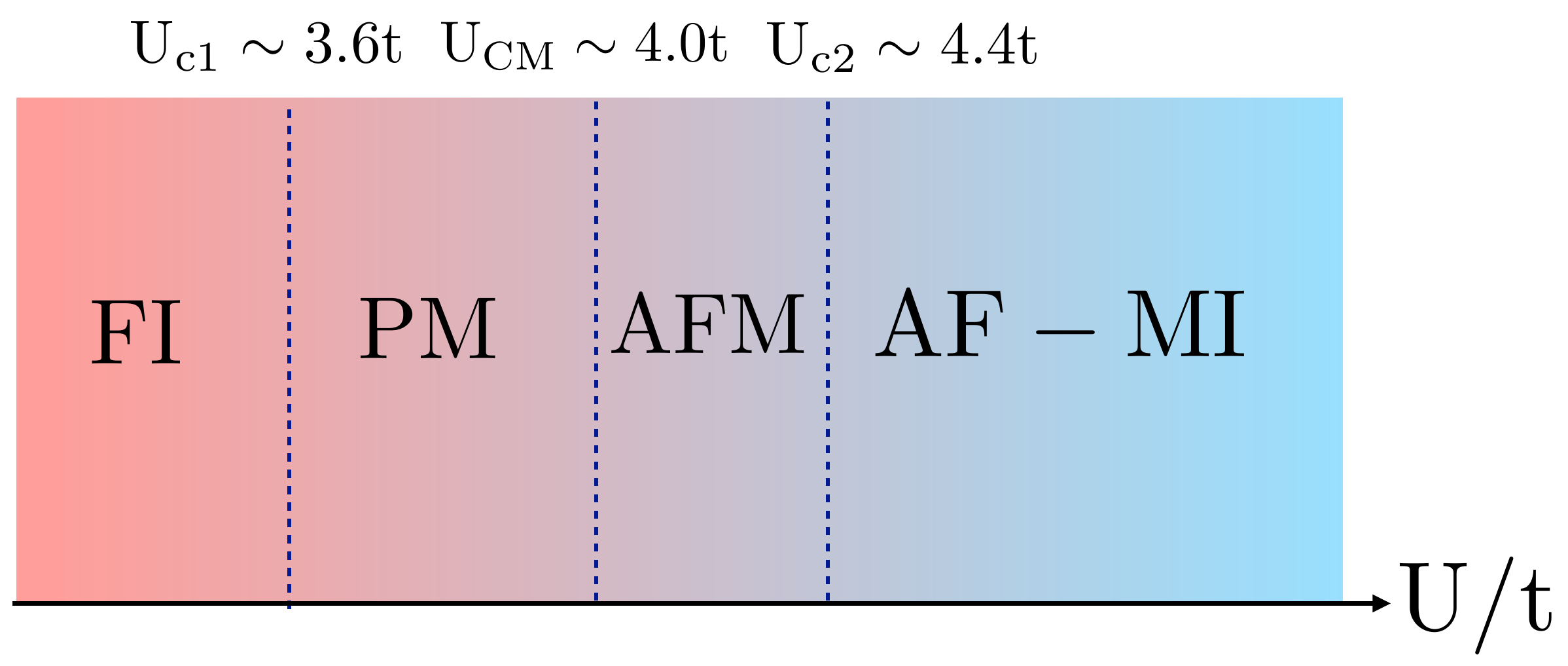}
\caption{Low temperature phase diagram (at $T=0.01t$) of the KHM at half filling mapping out 
the various thermodynamic phases as: (i) Flat band Insulator (FI), (ii) Paramagnetic 
Metal (PM), (iii) Antiferromagnetic Metal (AFM) and (iv) Antiferromagnetic Mott Insulator 
(AF-MI).}
\label{fig1}
\end{center}
\end{figure}

\textit{Phase diagram and critical points:}
We begin by presenting the low temperature phases of KHM at $T=0.01t$ which 
is $1/100^{th}$ of the reference energy scale,  in Fig. \ref{fig1}. Contrary to the existing consensus of a metal, 
we report a magnetically disordered flat band localized insulator in the weak coupling regime.
This NFL flat band localized insulator (FI) crosses over to a paramagnetic 
metal (PM) across the first critical point $U_{c1} \sim 3.6t$. 
The second critical point of this model marks the onset of the magnetic correlations at $U_{CM} \sim 4t$. The regime 
$U_{CM} < U \le U_{c2}$ hosts a metal, characterized by a (quasi) long ranged antiferromagnetic (AFM)
``precursor Coulomb phase'' with the $\sqrt{3} \times \sqrt{3}$ magnetic order carrying weights at the symmetry points of the extended 
Brillouin zone of the Kagome lattice. Further increase in $U$ leads the system through a first order MIT to an antiferromagnetic 
Mott insulator (AF-MI) across the third critical point $U_{c2} \sim 4.4t$. The MIT is accompanied by a symmetry breaking magnetic 
transition giving way to a Coulomb phase, quantified by a $\sqrt{3} \times \sqrt{3}$ magnetic order accompanied 
by the pinch points, in the static magnetic structure factor \cite{udagawa_prb2018,zhitomirsky_prb2008,shanon_prb2018,iqbal_prr2023}. 
Note that the MIT (at $U_{c2}$) and the onset of magnetic correlations (at $U_{CM}$) are well separated, in agreement 
with other magnetically frustrated systems such as, triangular, checkerboard lattice etc. \cite{nyayabanta_jpcm2017,mk_prb2022}. 

\begin{figure}
\begin{center}
\includegraphics[height=5.2cm,width=7.8cm,angle=0]{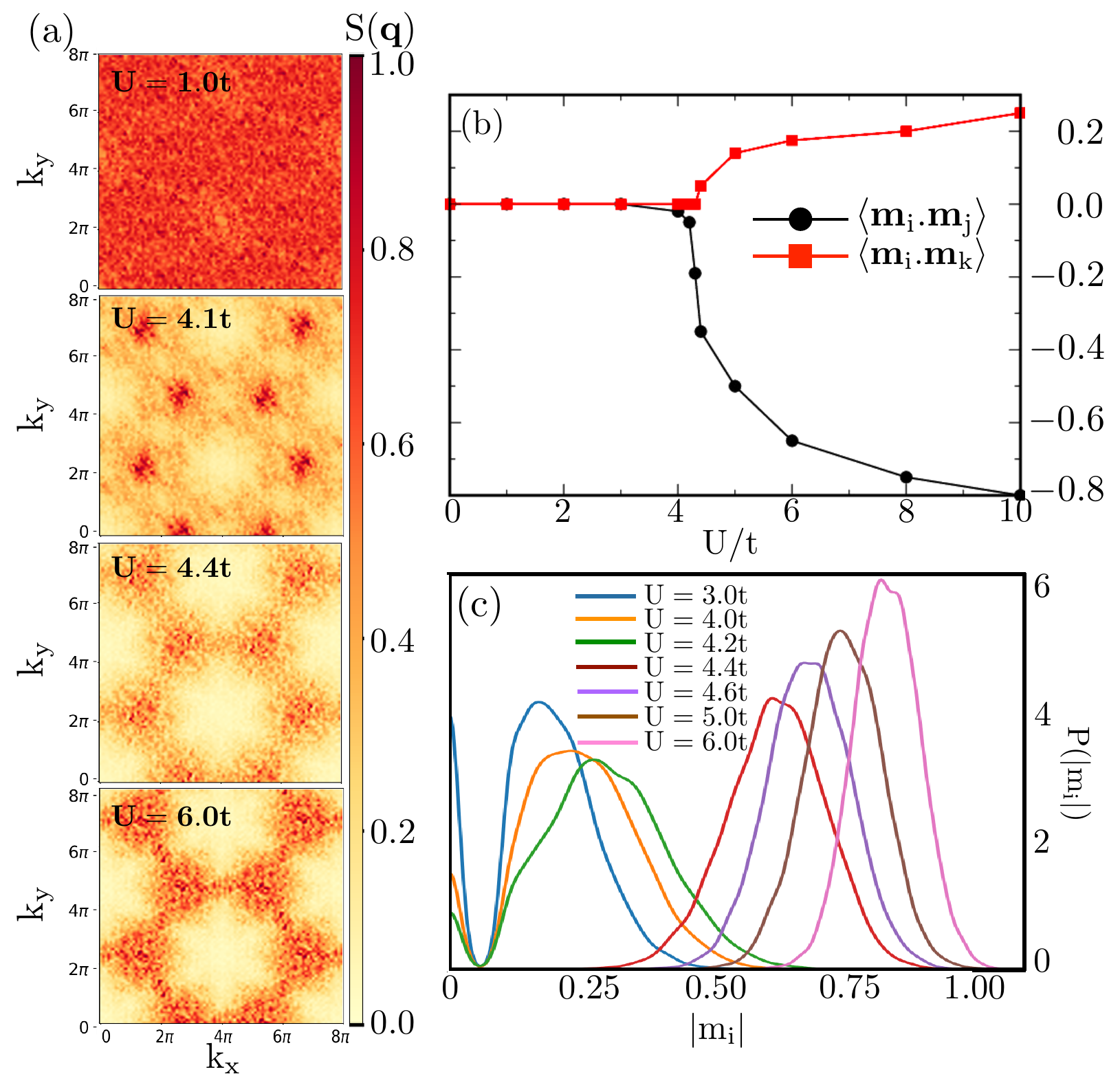}
\caption{Low temperature magnetic correlations at $T=0.01t$. (a) Magnetic structure factor ($S({\bf q})$) 
maps at the representative interactions: $U=t$, $U=4.1t$, $U=4.4t$ and $U=6.0t$. (b) Interaction dependent 
nearest neighbor ($\langle {\bf m}_{i}.{\bf m}_{j}\rangle$) (antiferromagnetic) and next nearest neighbor 
($\langle {\bf m}_{i}.{\bf m}_{k}\rangle$) (ferromagnetic) correlations. (c) Distribution 
of local magnetic moments ($P(\vert m_{i}\vert)$) at selected interactions.}
\label{fig2}
\end{center}
\end{figure}

\textit{The magnetic correlations:}
The thermodynamic signatures corresponding to the local moment formation and magnetic correlations at $T=0.01t$ are 
shown next in Fig. \ref{fig2},  at the representative interactions ($U$). The static magnetic structure factor 
($S({\bf q})$) maps across the critical points (CP) are shown in Fig.\ref{fig2}(a). The FI and PM regimes do not 
order magnetically and the corresponding $S({\bf q})$ maps are essentially featureless. The symmetry breaking 
magnetic transition to the precursor Coulomb phase across $U_{CM} \sim 4t$ is typified by the prominent and isolated 
$S({\bf q})$ peaks,  as shown at $U = 4.1t$. The $S({\bf q})$ maps shown at $U \sim 4.4t$ quantifies the second symmetry 
breaking magnetic transition to the Coulomb phase in the AF-MI regime. Interaction strengthens the Coulomb phase,  pushing 
the system towards the Heisenberg limit and rendering prominence to the pinch points as shown via the $S({\bf q})$ maps at $U=6t$.  
The pinch-points are well investigated both analytically and numerically in the context of classical nearest neighbor Heisenberg model 
on the Kagome lattice and are reported to arise from the flat band contributions \cite{udagawa_prb2018,zhitomirsky_prb2008,shanon_prb2018,iqbal_prr2023}. 

The real space local moment correlations between the nearest ($\langle {\bf m}_{i}.{\bf m}_{j}\rangle$) and 
the next nearest ($\langle {\bf m}_{i}.{\bf m}_{k}\rangle$) neighbors across the CPs are shown next in Fig.\ref{fig2}(b), 
wherein positive sign represents a ferromagnetic correlation while the negative sign indicates an antiferromagnetic  
correlation between the local moments. Magnetic correlations monotonically enhances with increasing interaction for 
$U \ge U_{CM}$ such that, the system develops a (quasi) long range antiferromagnetic order between the nearest neighbors, 
while a ferromagnetic order is dominant between the next nearest neighbors, in agreement with the 
results obtained based on the DMFT, D$\Gamma$A, DQMC and DCA calculations at $T \neq 0$ 
\cite{janson_prb2021}. The magnetic order saturates for $U \gg U_{c2}$ as indicated by their near independence 
of $U$.  

Insights into the behavior of ${\bf m}_{i}$ are obtained in terms of their distribution $P(\vert {\bf m}_{i}\vert)$, 
which we show next in Fig. \ref{fig2}(c). The distributions are quantified in terms of the mean amplitude of the local moments  
$\bar m$, the magnitude of $P(\vert {\bf m}_{i}\vert)$ and its width. The weak interaction regime $U \le U_{c1}$ is characterized 
by a bimodal distribution with the peaks centered around a small $\bar m \neq 0$ and at $\bar m =0$, indicating a spatially 
fragmented system with randomly oriented ${\bf m}_{i}$'s. Interaction progressively shifts the weight towards $\bar m \neq 0$ 
while at the same time suppressing the peak at $\bar m =0$,  the distribution however continues to remain broad, indicating a 
relatively ``weak'' magnetic order in the regime $U_{CM} < U \le U_{c2}$.  For $U > U_{c2}$, interaction has two-fold effects on 
$P(\vert {\bf m}_{i}\vert)$ as, $(i)$ the peak at $\bar m = 0$ is now completely suppressed and $(ii)$ the $\bar m \neq 0$ peak 
is robust, unimodal, sharp and monotonically shifts toward larger $\bar m$ values, implying an 
underlying spatially uniform Coulomb phase with (quasi) long range magnetic order.

\begin{figure}
\begin{center}
\includegraphics[height=6.5cm,width=8.2cm,angle=0]{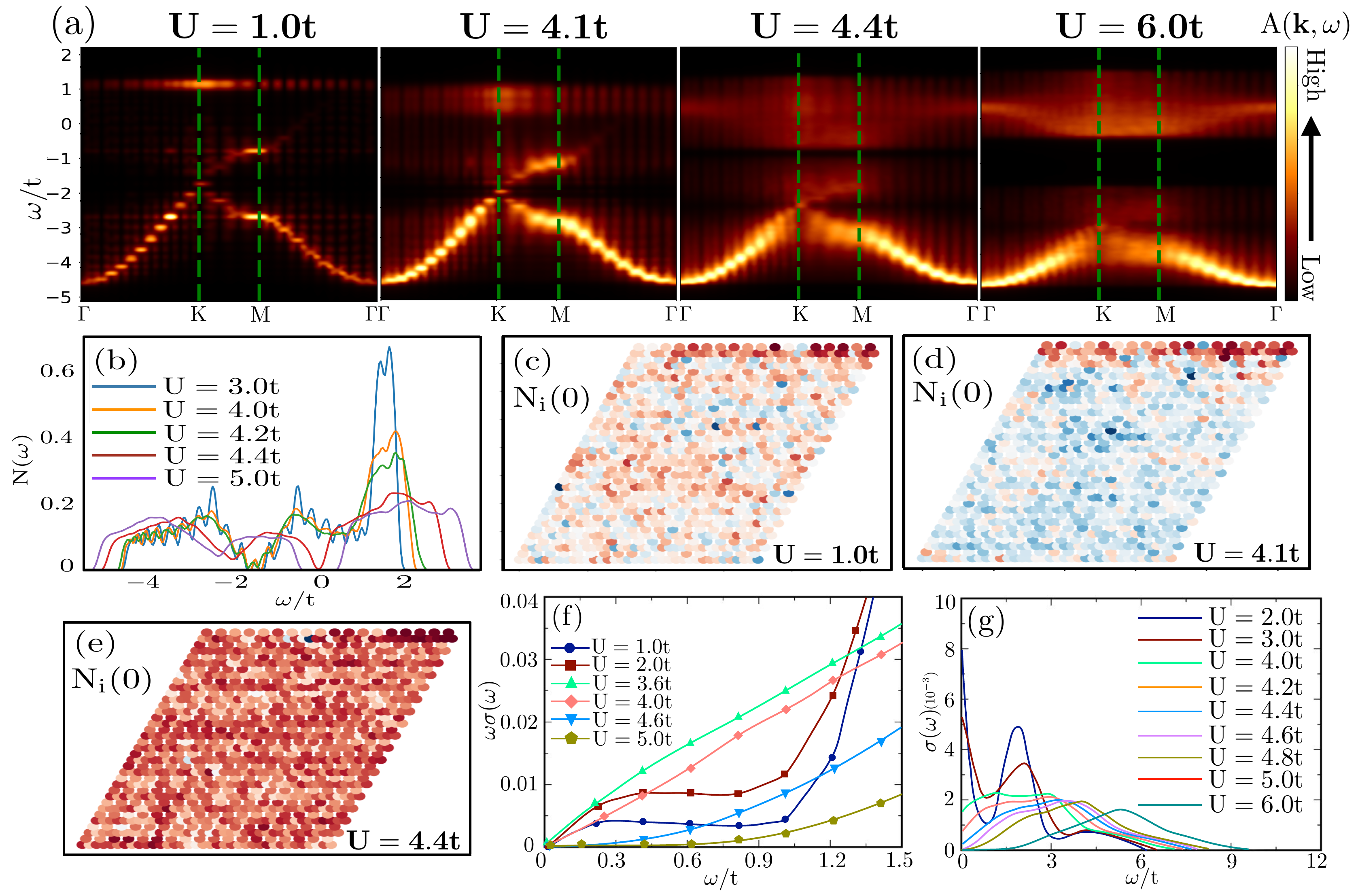}
\caption{Low temperature spectroscopic signatures at $T=0.01t$. (a) Spectral function (A(${\bf k}, \omega$)) maps 
showing the electronic dispersion at selected interactions. (b) Single particle DOS at selected interactions. 
(c)-(e) Maps corresponding to the LDOS at the Fermi level in the FI, AFM and AF-MI regimes. The colorbar 
(red:- low, blue:- high) shows the magnitude of the spectral weight at the Fermi level (see text for details).  
(f) Normalized optical conductivity ($\omega \sigma(\omega)$) at selected interactions. (g) $\omega \sigma(\omega)$ 
in the extended energy regime showing the Displaced Drude peak, (DDP).}
\label{fig3}
\end{center}
\end{figure}

\textit{Spectroscopic signatures:}
While the $S({\bf q})$ and the real space magnetic correlators quantify the magnetic phase transitions, 
the FI-PM crossover and the MIT across $U_{c1}$ and $U_{c2}$, respectively are determined based on the spectroscopic 
($A({\bf k}, \omega)$, $N(\omega)$, $N_{i}(0)$) and the transport ($\sigma(\omega)$) signatures of the system. We present 
the interaction tuned band structure renormalization of the system in terms of the spectral function $A({\bf k}, \omega)$, in 
Fig.\ref{fig3}(a). The dispersion spectra at $U=t$ is reminiscent of the non-interacting limit with a dominant high energy 
localized flat band and two dispersive bands with the Dirac point at the ${\bf K}$-point. Interaction alters 
this picture by,  $(i)$ broadening the bands and $(ii)$ increasing the overlap between the dispersive and flat bands at the 
$\Gamma$-point, in agreement with the high temperature DMFT, D$\Gamma$A and DCA calculations on 
the KHM \cite{janson_prb2021}. The $A({\bf k}, \omega)$ map at $U=4.4t$ shows that the Mott gap opens up at the Fermi 
level and progressively gets robust with interaction (see $U=6t$).

Fig.\ref{fig3}(b) shows the low temperature single particle DOS ($N(\omega)$) across the CPs. The regime 
$U < U_{CM}$ is gapless with a robust flat band at $\omega \sim 2t$. Interaction suppresses the flat band 
and the precursor Coulomb phase is typified by a soft gap in $N(\omega)$,  centered around the Fermi level. 
The MIT and the AF-MI phase for $U > U_{c2}$ is marked by the complete suppression of the high energy flat 
band and a robust Mott gap.   

We provide an insight to the evolution of the energy states quantifying the insulator-metal-insulator transitions 
across the CPs via the maps of the LDOS at the Fermi level ($N_{i}(0)$), shown in Fig.\ref{fig3}(c)-(e). The color intensity 
indicates the magnitude (red: low, blue: high) of the spectral weight. The FI at weak interactions ($U=t$) is gapless with a 
moderately suppressed LDOS corresponding to the flat band induced localization. Increasing interaction leads to the delocalization 
of the energy states and the gapless AFM phase at $U=4.1t$ with extended energy states host a large LDOS at the Fermi level. 
The Mott gap opening up at $U=4.4t$ strongly suppresses the LDOS and the same effectively vanishes deep inside the AF-MI regime.
  
\textit{Optical transport:}
Transport signatures probing the metal-insulator transition are shown next in Fig.\ref{fig3}(f) wherein 
we present the normalized optical conductivity ($\sigma(\omega)$) of the KHM across the CPs. 
For $U \le U_{c1}$ the low energy behavior of $\sigma(\omega)$ provides unambiguous signature of the flat band 
localized insulator with,  $\omega \sigma(\omega) \rightarrow 0$ {\it nonlinearly} as $\omega \rightarrow 0$. The FI 
is a NFL state, for which the Fermi liquid theory breaks down. Based on DMFT calculations, similar flat band localized 
NFL insulator has been reported to be stabilized in a Lieb lattice in the weak coupling regime of $U \sim t$ \cite{torma_prbl2021}. 
Increase in $U$ enhances the optical conductivity as the system inches towards the FI-PM crossover at $U_{c1}$ and the regime 
$U_{c1} < U \le U_{c2}$ is characterized by a {\it linear} $\omega \sigma(\omega) \rightarrow 0$ as $\omega \rightarrow 0$, 
as is expected from a metal. The $\sigma(\omega)$ drops with further increase in $U/t$ as the $\omega \rightarrow 0$ regime in the AF-MI 
phase hosts a Mott gap. 

Further analysis of the NFL signatures in the $U \lesssim U_{c2}$ regime is carried out and the results are 
encapsulated in Fig.\ref{fig3}(g) wherein the $\sigma(\omega)$ is presented over an extended energy range. 
The weak and intermediate interaction regimes are characterized by a high energy Displaced Drude peak (DDP) 
along with $\sigma(\omega) \neq 0$ in the low energy regime, indicating the underlying state to be a NFL. DDP has 
been extensively investigated as an unambiguous signature of NFL in the context of molecular charge transfer salts 
\cite{fratini_natcom2021}, organic Mott insulators \cite{kanoda_prl2020,itou_prl2020}, 2D transition metal dichalcogenides \cite{trivedi_prl2014,mak_nature2021,pasupathy_nature2021} etc., undergoing disorder or band width tuned MIT. 

Fig.\ref{fig3}(g) shows that for $U \le U_{c1}$ the DDP shifts towards high energies, while the $\omega \rightarrow 0$ 
peak undergoes progressive suppression with $U/t$, expected from an insulator.  In the regime $U_{c1} < U \le U_{c2}$,  
 $\sigma(\omega)$ is characterized by a broad DDP which progressively shift towards the low energies and $\sigma(\omega)$ 
has a finite weight at $\omega \rightarrow 0$ indicating a metallic state. The strong coupling regime of $U \ge U_{c2}$ is once 
again characterized by a high energy DDP along with an optical gap at $\omega \rightarrow 0$ corresponding to the Mott insulating 
phase. Increasing $U$ progressively shifts the DDP to higher energies as the system goes deep inside the Mott insulating regime.     
\begin{figure}
\begin{center}
\includegraphics[height=5.5cm,width=8.4cm,angle=0]{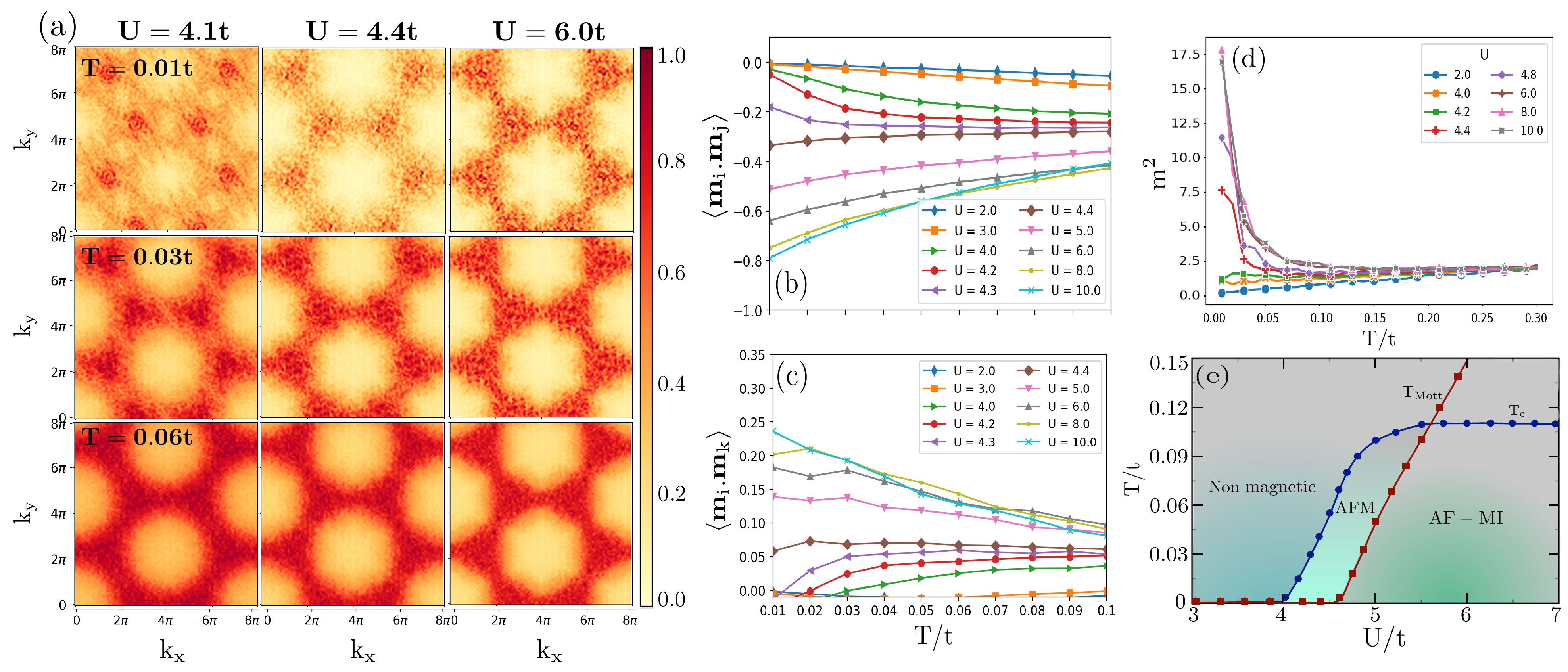}
\caption{Thermal evolution of the magnetic correlations. (a) Temperature dependence (along the columns) of 
$S({\bf q})$ at $U=4.1t$, $4.4t$ and $6.0t$. The high temperature $S({\bf q})$ maps out the Brillouin zone of the 
non interacting Kagome lattice. (b)-(c) Temperature dependence of the nearest and the next nearest neighbor 
magnetic correlations, respectively.  (d) Thermal evolution of the magnetic order parameter $m^{2}$ at 
selected interactions. The point of inflection of each curve quantifies the corresponding $T_{c}$. (e) Thermal phase 
diagram showing $T_{c}$ and $T_{Mott}$,  quantifying the loss of (quasi) long range magnetic order and spectral 
gap closure, respectively.}
\label{fig4}
\end{center}
\end{figure}

\textit{Thermal scales and magnetic correlations:}
Having established the low temperature phases we now focus on the impact of the thermal fluctuations on 
the various phases of this model. For the same we show the temperature dependence of 
$S({\bf q})$ at interactions representative of the precursor Coulomb and Coulomb phases, in 
Fig.\ref{fig4}(a). Thermal fluctuations destroy the (quasi) long range magnetic order and the corresponding 
$S({\bf q})$ progressively broadens indicating that the high temperature phase is dominated by short range 
correlations between the local moments. At still higher temperatures magnetic correlations are completely 
destroyed and the $S({\bf q})$ essentially maps out the non-interacting Brillouin zone of the Kagome lattice. 
Deep in the Mott regime (represented by $U=6t$) the magnetic correlations survive upto higher temperatures. 
The corresponding $S({\bf q})$ exhibits reminiscence of the low temperature pinch points.  

Next, the thermal evolution of the nearest ($\langle {\bf m}_{i}.{\bf m}_{j}\rangle$) and next nearest 
($\langle {\bf m}_{i}.{\bf m}_{k}\rangle$) neighbor magnetic correlations are presented in Fig.\ref{fig4}(b) and 
Fig.\ref{fig4}(c), respectively. For $U < U_{CM}$ the system doesn't order magnetically at low temperatures, however, 
the high temperature regime hosts weak but finite magnetic correlations arising out of the fluctuating local moments. 
Both the nearest and the next nearest neighbor correlations for $U \ge U_{CM}$ gets progressively suppressed via 
thermal fluctuations induced loss of the (quasi) long range magnetic order. We define a magnetic order parameter as 
$m^{2}$ (see SM) and track its thermal evolution to determine the magnetic transition temperature $T_{c}$ of the system, 
as shown in Fig.\ref{fig4}(d). The point of inflection of each $U$ curve quantifies the corresponding $T_{c}$. The thermal 
scale $T_{c}$, as determined is shown in Fig.\ref{fig4}(e). For $U> U_{CM}$, the $T_{c}$ is proportional to $U/t$, increases 
monotonically in the AFM regime and largely saturates in the AF-MI regime. Along with the $T_{c}$, an additional scale $T_{Mott}$ 
tracks the thermal evolution of the Mott gap at the Fermi level. The $T \neq 0$ thermodynamic phases are thus demarcated as 
non-magnetic, AFM and AF-MI, as shown in Fig.\ref{fig4}(e). The high temperature regime $T > T_{c}$ is dictated largely by the 
short range correlated fluctuating local moments. 

\begin{figure}
\begin{center}
\includegraphics[height=6.0cm,width=8.4cm,angle=0]{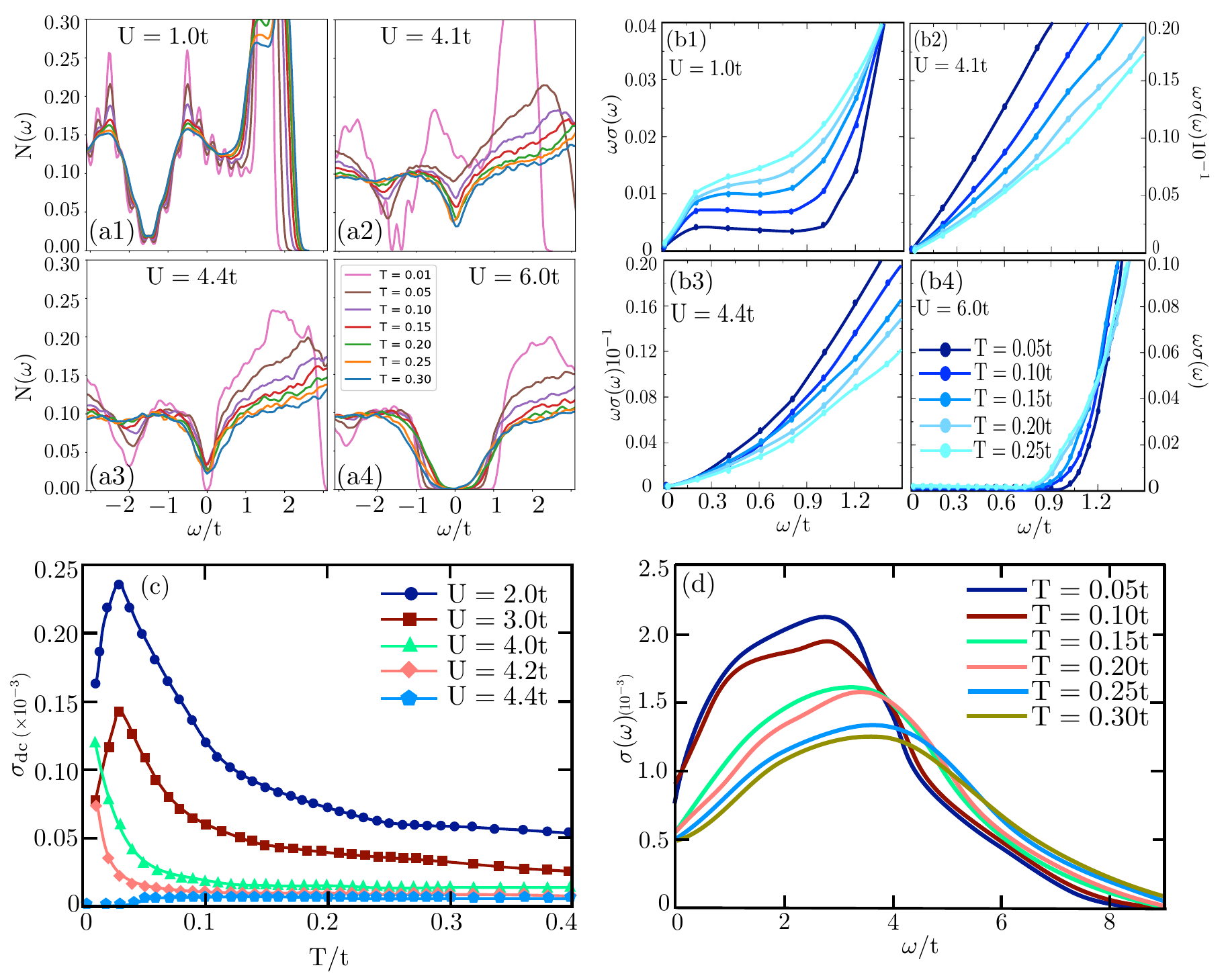}
\caption{Thermal evolution of the spectroscopic signatures. Temperature dependence of the single particle DOS 
(a1-a4) and normalized optical conductivity $\omega\sigma(\omega)$ (b1-b4), at the representative interactions. 
(c) Temperature dependence of the dc-conductivity ($\sigma_{dc}$) at selected interactions, with $d\sigma_{dc}/dT > 0$ 
corresponding to an insulating and $d\sigma_{dc}/dT < 0$ a metallic phase. (d) Thermal evolution of $\sigma(\omega)$ 
across the extended energy regime at the selected interaction of $U=4.2t$, highlighting the underlying NFL state.}
\label{fig5}
\end{center}
\end{figure}

\textit{Non-Fermi liquid signatures and quasiparticle transport:}
We next highlight the impact of thermal fluctuations on the spectroscopic and transport properties of the system, 
in terms of $N(\omega)$ and $\sigma(\omega)$, in Fig.\ref{fig5}(a) and \ref{fig5}(b), respectively. In the FI-NFL regime 
the single particle DOS is gapless at all temperatures,  with a finite spectral weight at the Fermi level and a robust flat 
band at $\omega \sim 2t$, which undergoes progressive suppression with temperature. The gapless NFL metal in the 
PM regime develops a dip in the single particle DOS at the Fermi level at high temperatures owing to the thermal fluctuations 
induced short range correlations.  A soft gap opens up at the Fermi level at the MIT boundary ($U=4.4t$) which rapidly fills up 
via accumulation of spectral weight, resulting in a V-shaped spectra akin to that observed in the pseudogap phase  of the high 
temperature superconductors, establishing a NFL state. The Mott gap at the Fermi level is largely immune to fluctuations as 
shown at $U=6t$. 
 
Fig.\ref{fig5}(b) shows the thermal evolution of $\sigma(\omega)$ in the low energy regime. Thermal fluctuations 
induced randomness of the local moments progressively destroys the insulating behavior of the FI-NFL regime 
($U = t$). In a similar spirit the metallicity of the AFM regime is suppressed with temperature and $\sigma(\omega)$  
reduces (see $U=4.1t$).  At $U=4.4t$,  $\omega\sigma(\omega)$ weakly deviates from the metallic behavior indicating 
the onset of MIT. Thermal fluctuation suppresses the metallicity as observed from the drop in $\sigma(\omega)$. 
The strong interaction regime of AF-MI hosts a gapped optical spectra with $\sigma(\omega)$ being largely immune 
to the thermal fluctuations.  

For a strongly correlated system the Fermi liquid behavior or the deviation from it is quantified based on three 
important signatures in its transport properties viz. $(i)$ the low energy dependence of $\sigma(\omega)$, 
$(ii)$ deviation of electrical resistivity from $\rho_{xx} \propto T^{2}$ behavior and $(iii)$ DDP at high energies 
in $\sigma(\omega)$ \cite{han_rmp2003,takagi_philmag2004,georges_prl2013}. Fig.\ref{fig5}(b) discusses 
$(i)$ and in order to understand $(ii)$ and $(iii)$ we present Fig.\ref{fig5}(c) and Fig.\ref{fig5}(d) next. The 
$\omega \rightarrow 0$ limit of $\sigma(\omega)$ quantifies the dc-conductivity ($\sigma_{dc}$), which we show 
next in Fig.\ref{fig5}(c). We distinguish the metallic and insulating phases based on the temperature dependence 
of $\sigma_{dc}$ such that,  $d\sigma_{dc}/dT < 0$  corresponds to a metal, while an insulator is realized for 
$d\sigma_{dc}/dT > 0$. Fig.\ref{fig5}(c) shows that the weak coupling regime at low temperature is characterized 
by $d\sigma_{dc}/dT > 0$, in agreement with the underlying FI phase. For the metallic phase at the intermediate 
interactions $d\sigma_{dc}/dT < 0$, at low temperatures. The AF-MI at strong coupling is typified by 
an optical gap at low temperatures. Independent of the choice of $U$,  for the high temperature phase 
$d\sigma_{dc}/dT < 0$, in agreement with the thermal fluctuations induced enhanced metallicity. The non 
monotonic temperature dependence of $\sigma_{dc}$,  as observed from Fig.\ref{fig5}(c) indicates deviation 
of the corresponding resistivity ($\rho_{xx} = 1/\sigma_{dc}$) from $\rho_{xx} \propto T^{2}$ behavior, thereby 
affirming an underlying NFL state. It must be noted that the breakdown of the Fermi liquid description of the 
system doesn't correspond to the breakdown of its quasiparticle description.  The physics of the NFL metal 
can be captured in terms of the {\it resilient} quasiparticles which survives as the relevant low energy 
excitations even when the Fermi liquid description of the system breaks down \cite{han_rmp2003,takagi_philmag2004,georges_prl2013}. 

The temperature dependence of $\sigma(\omega)$ in the high energy regime is shown in Fig.\ref{fig5}(d) for a selected 
interaction of $U=4.2t$. The low temperature NFL metal is characterized by $\sigma(\omega) \neq 0$ as $\omega \rightarrow 0$ 
and a high energy DDP. With increasing temperature the DDP marginally shifts towards higher energies indicating weak suppression 
in the metallicity.  The breakdown of the resilient quasiparticle description is signaled by,  $(i)$ a broadening of the DDP and its progressive 
shift towards higher energies and $(ii)$ the loss of the isosbestic crossing point of the $\sigma(\omega)$ curves. While $(i)$ can be 
readily observed from Fig.\ref{fig5}(d), a closer investigation of $\sigma(\omega)$ at high temperature reveals that for $T \gtrsim 0.2t$ 
the isosbestic crossing point of the curves is lost, indicating the gradual decay of the quasiparticles as the system crosses over to a bad 
metallic phase where the Fermi liquid theory breaks down \cite{han_rmp2003,takagi_philmag2004,georges_prl2013}. 

\textit{Discussion and conclusions:}
In this work we employed a non perturbative numerical approach to access the physics of KHM at half filling, away 
from the strong coupling regime. While SPA which doesn't include the quantum fluctuations can't capture the true 
ground state of KHM, it provides important insight to the low temperature physics of this system, which remains inaccessible to most of the existing numerical techniques, owing to severe fermionic sign problem and strong finite 
size effects. 

One of the key results that we showed in this letter is the flat band localized gapless insulating phase deep in the weak 
coupling regime, in contrast to the existing consensus of a metallic phase.  This gapless insulating phase can't be 
discerned via the standard spectroscopic indicators such as, the single particle DOS and spectral function. The low 
energy dependence of the optical transport properties, computed here for the first time for the KHM however,  exhibits distinct non linearity,  establishing a NFL insulator. Our results significantly alters the existing understanding of the low temperature phase in the weak coupling regime of the KHM and exhibits that a NFL insulator-metal crossover takes 
place at $U_{c1} \sim 3.6t$. 

Our second key observation from this work concerns the MIT in the KHM,  stabilized at the other end 
of the interaction axis. Based on DQMC, D$\Gamma$A, CDMFT, VCA etc. the estimate of $U_{c}$ for the MIT 
ranges  between $\sim 5t$ to $\sim 11t$ and is reported to be sensitive to non local correlations \cite{tsunetsugu_prl2006,tsunetsugu_jpcm2007,asano_prb2016,hatsugai_prl2019,ohashi_prl2006,maekawa_prl2005,janson_prb2021,paiva_prb2023,held_rmp2018,held_prb2021,toschi_prb2016,asano_prb2016}. The underlying magnetic correlation is largely agreed upon to be a $\sqrt{3} \times \sqrt{3}$ phase with the characteristic static structure factor 
maps having maxima at the ${\bf K}$-points of the extended Brillouin zone. Our results give the estimate of 
$U_{c2} \sim 4.4t$ for the first order MIT. The magnetic state is $\sqrt{3} \times \sqrt{3}$ correlated in agreement 
with the existing literature. However, rather than exhibiting only high intensity spots at the ${\bf K}$-points   
of the extended Brillouin zone, our $S({\bf q})$ maps provide unambiguous signature of the pinch points, the salient 
signature of magnetic frustration. The $\sqrt{3} \times \sqrt{3}$  magnetic correlation in conjunction with the pinch points 
correspond to the low temperature Coulomb phase of the $S=1/2$ Kagome Heisenberg model  \cite{udagawa_prb2018,zhitomirsky_prb2008,shanon_prb2018,iqbal_prr2023}. Our results therefore 
exhibits for the first time that the strong coupling regime of KHM maps on to the $S=1/2$ Kagome Heisenberg model. For $U \gg U_{c2}$ this mapping gets progressively stronger as the system goes deep in the Heisenberg limit (see SM). Note that the Coulomb phase captured by SPA is a low temperature phase of the strong coupling KHM and not its true ground state. Within the purview of SPA the system remains arrested in the Coulomb phase at still lower temperatures, as the neglect of quantum fluctuations in this scheme restricts the system from reaching the true ground state of coplanar magnetic order via thermal order by disorder mechanism \cite{udagawa_prb2018}. 

The intermediate interaction regime $U_{c1} < U \le U_{c2}$ is a NFL metal which can further be demarcated by the 
$U_{CM} \sim 4.0t$ into PM and AFM.  While a stable metallic phase in the intermediate interaction regime of KHM 
was reported based on DQMC, DMRG, D$\Gamma$A studies \cite{tsunetsugu_prl2006,tsunetsugu_jpcm2007,asano_prb2016,hatsugai_prl2019,ohashi_prl2006,maekawa_prl2005,janson_prb2021,paiva_prb2023,held_rmp2018,held_prb2021,toschi_prb2016,asano_prb2016}, to the best of our knowledge this is the first time that the underlying 
magnetic correlations are discussed. The AFM phase ($U_{CM} < U \le U_{c2}$) referred to as the precursor Coulomb phase 
in this work and characterized by isolated $S({\bf q})$ peaks at the ${\bf K}$-points of the extended Brillouin zone of the Kagome 
lattice is distinct from the Coulomb phase observed for $U \ge U_{c2}$, in its lack of the pinch points. We note that while the magnetic 
phases are verified to be stable against finite system size effects, quantum fluctuations can reduce their regime of stability or destabilize 
them. Unfortunately for the existing DQMC or DMFT based studies this regime of temperature remains inaccessible due to severe fermionic 
sign problem. A recent DMRG based calculation however, reported two different symmetry broken magnetic states as a function 
of $U$, in the KHM at half filling \cite{zhu_prbl2021}, akin to our observations presented in this letter. 

Finally, we comment on the relevance of the results discussed in this letter from the experimental perspective. The 
recent interest on Kagome metal or KHM away from the strong coupling regime stems from the discovery of FeSn, Mn$_{3}$Sn and more recently the AV$_{3}$Sb$_{5}$ family \cite{higo_nature2015,parkin_sciadv2016,shin_natmat2017,otani_nature2019,behnia_natcom2019,hess_prb2019,wills_jpcm2009,wills_jpcm2011,zhang_advmat2017,checkelsky_nature2018,wang_nature2018,zhang_prl2018,wang_apl2019,tranquada_prl2019,kondo_prb2020,chen_natphys2018,lei_natcom2018,hasan_natphys2019,chen_science2019,shen_apl2019,analytis_natcom2020,madhavan_natcom2021,ochiai_prb2018,ochiai_jspj2019,checkelsky_apl2019,ghimire_natmat2019,zhang_prb2020,mcguire_prm2019,wilson_prm2021,wilson_prl2020}. The results of KHM at half filling however can't be directly corroborated to the experimental observations owing to the interplay of competing interaction scales, multiband nature of the real materials etc. except perhaps for FeSn,  whose band structure is akin to that of the non interacting 
single band Kagome lattice \cite{comin_natmat2019,zhang_prb2020}. Nevertheless, the single band KHM discussed 
here provides the vital entry point to this complex interplay of non local correlations, magnetic frustration, and flat 
electronic band induced localization effects, in addition to the impact of the thermal fluctuations. 

Along with the intriguing low temperature physics brought forth by the interplay of spin and charge degrees of freedom, 
we established based on the transport properties that the high temperature regime of this system is a NFL metal characterized 
by resilient quasiparticles which undergoes progressive degradation to a bad metallic phase. Normal state resistivity and optical 
conductivity measurements should bore out the signatures of this NFL metal.  The spectroscopic signatures as depicted in this 
work in terms of the spectral functions should be accessible to the angle resolved photo emission spectroscopy (ARPES) measurements, 
while neutron scattering experiments are suitable probe to the magnetic correlations. In short, more experimental results 
are awaited from the real Kagome materials with closer semblance to the single band KHM in terms of material parameters, band 
structure etc. and our results hold promise to initiate the same.

\textit{Acknowledgements:} MK would like to acknowledge the use of the high performance computing facility 
(AQUA) at the Indian Institute of Technology, Madras, India.

\section{Supplementary Material}

\textit{Model, method and indicators}

Our starting Hamiltonian is the repulsive Hubbard model on a Kagome lattice, defined as, 
\begin{eqnarray}
\hat{H} = -t \sum_{\langle ij  \rangle,\sigma}(\hat c^{\dagger}_{i,\sigma}\hat c_{j,\sigma} + h.c) - \mu \sum_{i,\sigma} \hat n_{i\sigma} \nonumber + U \sum_{i}\hat n_{i,\uparrow}\hat n_{i,\downarrow} \nonumber \\ 
\end{eqnarray}

\noindent here, $t_{ij}=t$ is the nearest neighbor hopping on a Kagome lattice and $t=1$ sets the reference 
energy scale of the problem. $U > 0$ is the on-site repulsive Hubbard interaction. We work at half filling and 
the chemical potential $\mu$ is adjusted to achieve the same. In order to make the model numerically tractable 
we decompose the interaction term using Hubbard Stratonovich (HS) decomposition \cite{hs1,hs2}.  
This results in the introduction of two (bosonic) auxiliary fields viz. a vector field ${\bf m}_{i}(\tau)$ and a scalar field 
$\phi_{i}(\tau)$, which couples to the spin and charge densities, respectively. The introduction of these auxiliary 
fields retain the spin rotation invariance and the Goldstone modes, and allows us to capture the Hartree-Fock theory 
at the saddle point, . In terms of the Grassmann fields $\psi_{i\sigma}(\tau)$, we write,

\begin{eqnarray}
\exp[U\sum_{i}\bar\psi_{i\uparrow}\psi_{i\uparrow}\bar\psi_{i\downarrow}\psi_{i\downarrow}] & = & \int {\bf \Pi}_{i}
\frac{d\phi_{i}d{\bf m}_{i}}{4\pi^{2}U}{\exp}[\frac{\phi_{i}^{2}}{U}+i\phi_{i}\rho_{i}+\frac{m_{i}^{2}}{U} 
\nonumber \\ && -2{\bf m}_{i}.{\bf s}_{i}]
\end{eqnarray}

where, the charge and spin densities are defined as, $\rho_{i} = \sum_{\sigma}\bar\psi_{i\sigma}\psi_{i\sigma}$ 
and ${\bf s}_{i}=(1/2)\sum_{a,b}\bar \psi_{ia}{\bf \sigma}_{ab}\psi_{ib}$, respectively. The corresponding 
partition function thus takes the form,

\begin{eqnarray}
{\cal Z} & = & \int {\bf \Pi}_{i}\frac{d\bar\psi_{i\sigma}d\psi_{i\sigma}d\phi_{i}d{\bf m}_{i}}{4\pi^{2}U}
\exp[-\int_{0}^{\beta}{\cal L}(\tau)]
\end{eqnarray}
where, the Lagrangian ${\cal L}$ is defined as,
\begin{eqnarray}
{\cal L}(\tau) & = & \sum_{i\sigma}\bar\psi_{i\sigma}(\tau)\partial_{\tau}\psi_{i\sigma}(\tau) + H_{0}(\tau) 
\nonumber \\ && +\sum_{i}[\frac{\phi_{i}(\tau)^{2}}{U}+(i\phi_{i}(\tau)-\mu)\rho_{i}(\tau)+
\frac{m_{i}(\tau)^{2}}{U} \nonumber \\ && -2{\bf m}_{i}(\tau).{\bf s}_{i}(\tau)]
\end{eqnarray}
where, $H_{0}(\tau)$ is the kinetic energy contribution. 
The $\psi$ integral is now quadratic but at the cost of an additional integration over
the fields ${\bf m}_{i}(\tau)$ and $\phi_{i}(\tau)$. The weight factor for the ${\bf m}_{i}$
and $\phi_{i}$ configurations can be determined by integrating out the $\psi$ and
$\bar \psi$;  and using these weighted configurations one goes back and computes
the fermionic properties. Formally,

{\begin{eqnarray}
{\cal Z} & = & \int {\cal D}{\bf m}{\cal D}{\phi}e^{-S_{eff}\{{\bf m},\phi\}}
\end{eqnarray}}
\begin{eqnarray}
S_{eff} & = & \log Det[{\cal G}^{-1}\{{\bf m},\phi\}] + \frac{\phi_{i}^{2}}{U} +
\frac{m_{i}^{2}}{U}
\end{eqnarray}
where, ${\cal G}$ is the electron Green's function in a $\{{\bf m}_{i},\phi_{i}\}$ background.

The weight factor for an arbitrary space-time configuration $\{{\bf m}_{i}(\tau), \phi_{i}(\tau)\}$ 
involves computation of the fermionic determinant in that background. The auxiliary field quantum Monte Carlo 
with static path approximation (SPA) retains the full spatial dependence
in ${\bf m}_{i}$ and $\phi_{i}$ but keeps only the zero mode of the Matsubara frequency 
($\Omega_{n}=0$). It thus includes classical fluctuations of arbitrary magnitudes but no quantum
($\Omega_{n} \neq 0$) fluctuations. 

Following the SPA approach, we freeze $\phi_{i}(\tau)$ to its saddle point value 
$\phi_{i}(\tau)=\langle n_{i} \rangle U/2$, where $\langle n_{i}\rangle$ is the fermionic 
number density. The resulting model can be thought 
of as fermions coupled to spatially fluctuating random background of classical field ${\bf m}_{i}$. With 
these approximations the effective Hamiltonian corresponds to a coupled spin-fermion model and reads as, 

\begin{eqnarray}
H_{eff} & = & \sum_{\langle ij\rangle, \sigma}t_{ij}[c_{i\sigma}^{\dagger}c_{j\sigma}+h.c.] 
-\tilde \mu \sum_{i\sigma} \hat n_{i\sigma} \nonumber \\ && - \frac{U}{2}\sum_{i}{\bf m}_{i}.{\bf \sigma}_{i} 
+ \frac{U}{4}\sum_{i}m_{i}^{2}
\end{eqnarray}
where, $\tilde \mu =\sum_{i}(\mu-\langle n_{i}\rangle U/2)$ and the last term of $H_{eff}$ corresponds to the 
stiffness cost associated with the now classical field 
${\bf m}_{i}$ and ${\bf \sigma}_{i}=\sum_{a,b}c_{ia}^{\dagger}{\bf \sigma}_{ab}c_{ib}={\bf s}_{i}$.

The random background configurations of $\{{\bf m}_{i}\}$ are generated numerically via Monte Carlo 
simulation and obey the Boltzmann distribution,
\begin{eqnarray}
P\{{\bf m}_{i}\} \propto Tr_{c,c^{\dagger}}e^{-\beta H_{eff}}
\end{eqnarray}
\begin{figure}
\begin{center}
\includegraphics[height=8cm,width=8cm,angle=0]{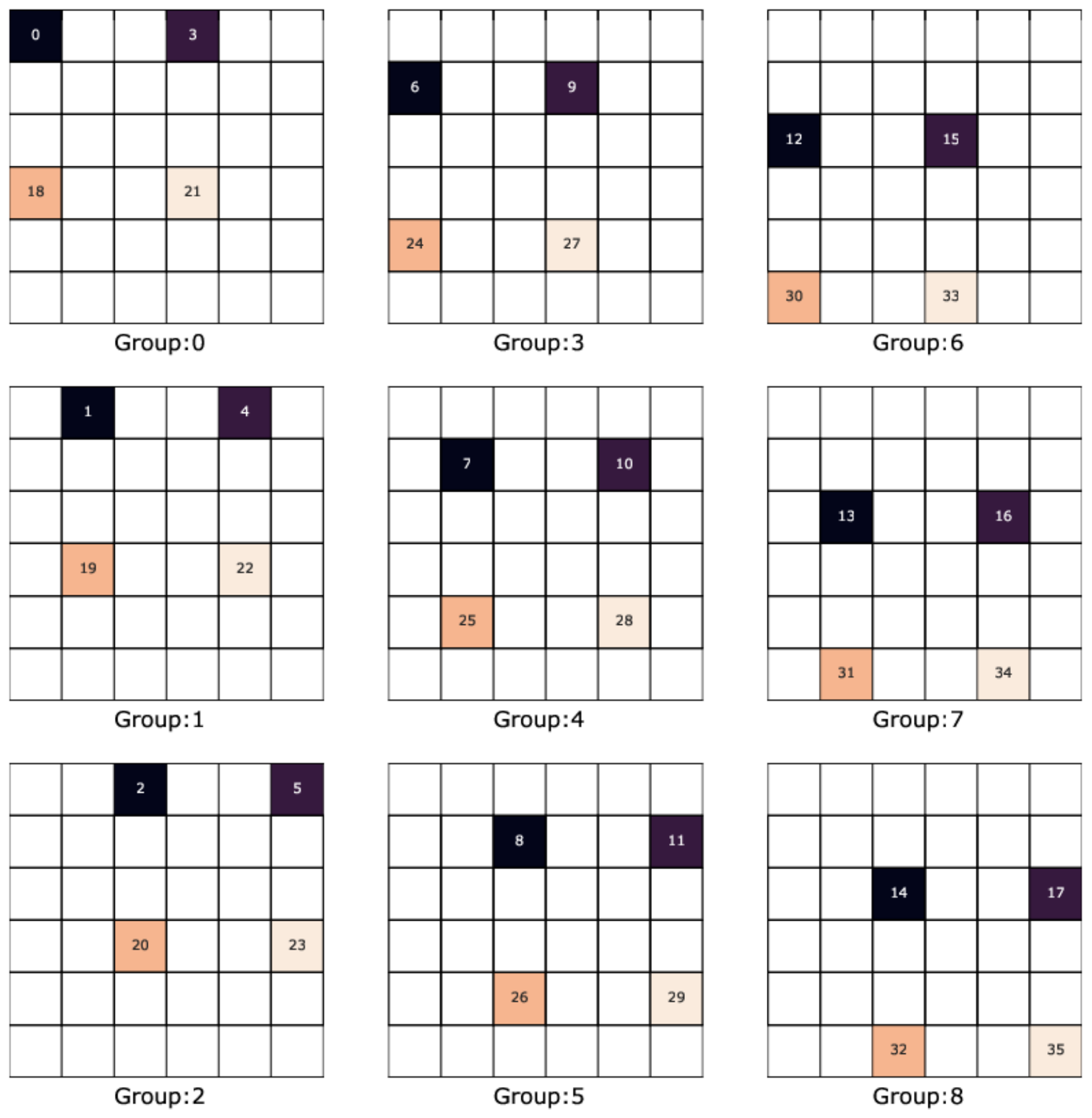}
\caption{PTCA scheme for a $6 \times 6$ lattice using a $4 \times 4$ cluster. One first needs to form groups of sites 
that can be updated simultaneously. For a given cluster size one can have different number of groups. For $4 \times 4$ 
cluster we have 9 such groups . Each group consists of sites that can be updated simultaneously since the cluster involving 
environment consisting of neighboring site is kept fixed. Once sites within one group is updated we move to update the sites 
in the next group. The next group consists of sites that were updated recently.}
\label{fig1_suppl}
\end{center}
\end{figure}

For large and random configurations the trace is computed numerically, wherein we diagonalize $H_{eff}$ for each attempted 
update of ${\bf m}_{i}$ and converge to the equilibrium configuration using Metropolis algorithm. Evidently, the process is numerically 
expensive. 

\textit{TCA and PTCA}

For a $N=3L^{2}$ site system, the dimensionality of hamiltonian is $2N\times 2N$. Each diagonalization step has 
a computational complexity of the order $\mathcal{O}(N^{3})$. Thus, increasing the system size by factor of 2 ($L \rightarrow 2L$) 
increases the computational complexity by 8 times. This is made worse by the fact that we have to update all $N$ sites to 
perform one Monte-Carlo step. And to obtain equilibrium configurations of the HS variables we perform $\mathcal{O}(10^{3})$ 
Monte-Carlo steps. Once equilibration is achieved we make measurements to mimic the ensemble averaging. Generally we 
make close to $\mathcal{O}(10^{3})$ measurements. Thus, direct application of this scheme in this form restrict us to smaller 
system sizes. To solve this problem a traveling cluster approximation (TCA) is used within the Monte-Carlo scheme. The scheme 
consists of diagonalizing a smaller cluster consisting of $M$ sites $(M \ll N)$. The center of the cluster has the site that we aim to 
update and the cluster consists of the neighboring sites. This scheme has been proved very effective to simulate larger system sizes 
\cite{mk_pra2016,mk_pra2018,nyayabanta_prb2016,nyayabanta_jpcm2017,mk_prb2022}. 
One of the advantage of the scheme is that the computational complexity associated with the diagonalization doesn't change 
with $N$ once we have fixed the number of sites in the cluster as $M$.

However, still performing simulations for bigger system sizes (large $N$) is difficult.  The TCA not only allows us to reduce 
the computational complexity from $\mathcal{O}(N^{3})$  to $\mathcal{O}(M^{3})$ but it also allows us to parallelize our code 
on multiple CPUs. Thus, allowing us to access much bigger system sizes. The approximation is called parallelized TCA (PTCA ) 
\cite{dagotto_pre2015}. The $N$ site update during each Monte-Carlo step can be split over $K$ CPU's. The finite size clusters 
used in updating the sites allows us to form groups of sites that can be updated in parallel (see Fig.\ref{fig1_suppl} for details). Once 
all these sites in one group are updated we use the updated sites to form clusters for sites in the second group. In our example 
(Fig.\ref{fig1_suppl}) we are forming clusters of $4 \times 4$. We simulate a system of linear dimension $L$ with a total number 
of unit-cell $N=L^{2}$ and the total number of sites being $3N$.

\textit{Indicators}

Once the equilibrium configurations of $\{{\bf m}_{i}\}$ are obtained the different phases are characterized based on the following fermionic 
correlation functions, 
\begin{figure*}
\begin{center}
\includegraphics[height=8cm,width=15cm,angle=0]{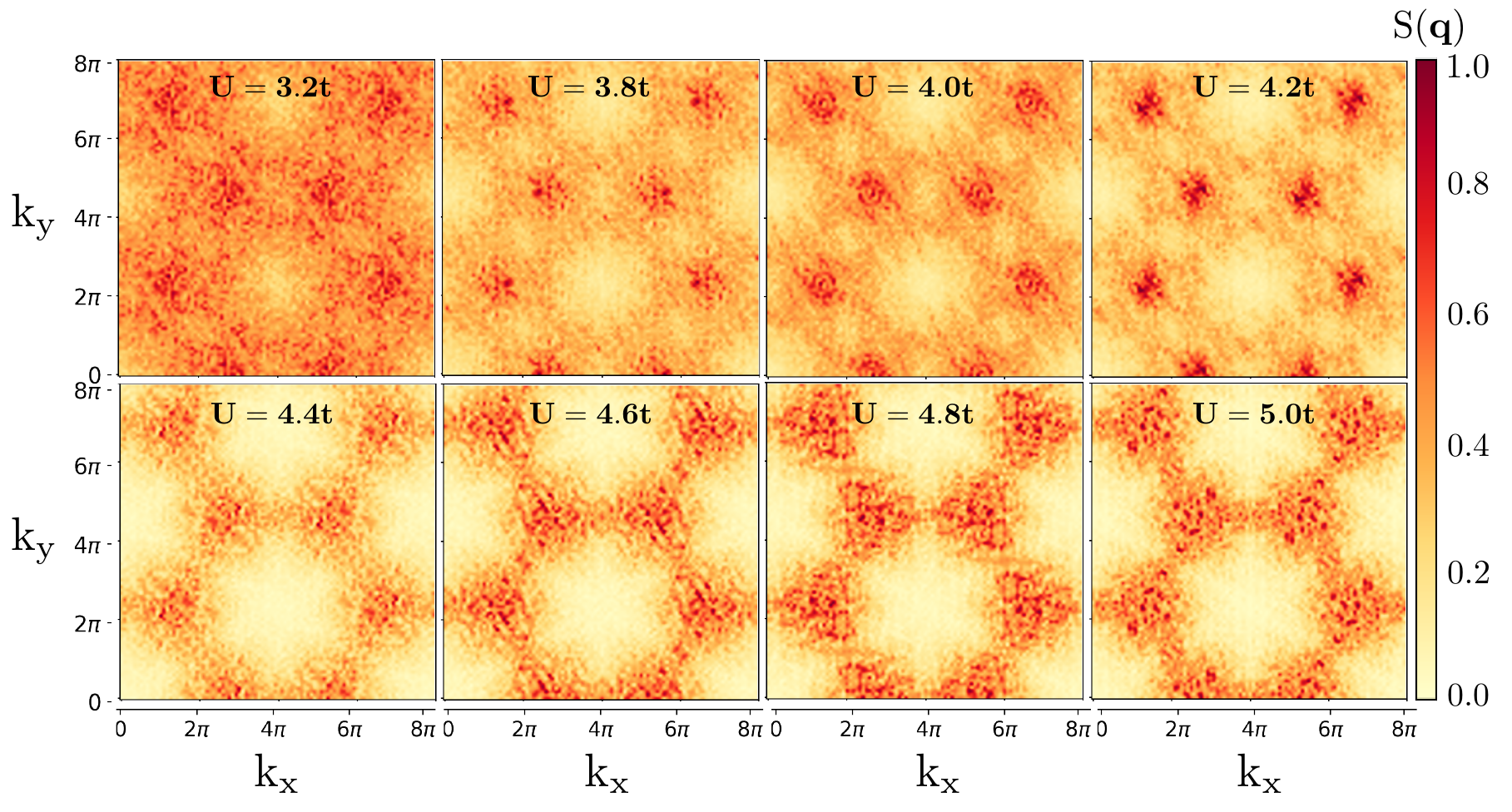}
\caption{Low temperature magnetic structure factor maps across the critical points. The precursor Coulomb phase ($U_{CM} < U \le U_{c2}$) is characterized by isolated $S({\bf q})$ peaks mapping out a $\sqrt{3} \times \sqrt{3}$ order. The AF-MI phase for $U \ge U_{c2}$ is quantified by pinch points accompanying the  $\sqrt{3} \times \sqrt{3}$ order, corresponding to the Coulomb phase.}
\label{fig2_suppl}
\end{center}
\end{figure*}

\begin{itemize}
\item{Magnetic structure factor,
\begin{eqnarray}
S({\bf q}) & = & \frac{1}{N^{2}}\sum_{ij}\langle {\bf m}_{i}.{\bf m}_{j}\rangle e^{i {\bf q}.({\bf r}_{i}-{\bf r}_{j})}
\end{eqnarray}
where, ${\bf q}$ corresponds to the magnetic ordering wave vector and $N$ is the number of lattice sites. 
$\langle ... \rangle$ corresponds to the Monte Carlo configurational average.
}
\item{Single particle density of states (DOS)
\begin{eqnarray}
N(\omega) & = & (1/N)\sum_{n}\langle \delta(\omega - \epsilon_{n})\rangle
\end{eqnarray}
where, $\epsilon_{n}$ are eigenvalues in a single equilibrium configuration. 
}
\item{Single particle local density of states (LDOS)
\begin{eqnarray}
N_{i}(\omega) & = & \sum_{n, \sigma}\langle \vert u^{i}_{n, \sigma}\vert^{2} \delta(\omega-\epsilon_{n})\rangle
\end{eqnarray}
where, $u^{i}_{n, \sigma}$ is the eigenvector corresponding to the eigenvalues $\epsilon_{n}$. 
}
\item{Optical conductivity, calculated using the Kubo formula, 
\begin{eqnarray}
\sigma(\omega) & = & \frac{\sigma_{0}}{N} \sum_{\alpha, \beta} \frac{f(\epsilon_{\alpha})-f(\epsilon_{\beta})}{\epsilon_{\beta}-\epsilon_{\alpha}} \vert \langle \alpha \vert J_{x} \vert \beta \rangle \vert^{2} \delta(\omega - (\epsilon_{\beta}-\epsilon_{\alpha})) \nonumber \\ 
\end{eqnarray}
where, the current operator $J_{x}$ is defined as, 
\begin{eqnarray}
J_{x} & = & -i\sum_{i, \sigma, \vec \delta} [{\vec \delta}t_{\vec \delta}c_{{\bf r}_{i}, \sigma}^{\dagger}c_{{\bf r}_{i}+\vec \delta, \sigma} - H. c.]
\end{eqnarray}
The dc conductivity ($\sigma_{dc}$) is the $\omega \rightarrow 0$ limit of $\sigma(\omega)$, $\sigma_{0}=\frac{\pi e^{2}}{\hbar}$ in 2D. $f(\epsilon_{\alpha})$ is the Fermi function, and $\epsilon_{\alpha}$ and $\vert \alpha\rangle$ are respectively the single particle eigenvalues and eigenvectors of $H_{eff}$ in a given background of $\{{\bf m}_{i}\}$.
}
\item{Spectral function, 
\begin{eqnarray}
A({\bf k}, \omega) & = & -(1/\pi){\mathrm{Im}}G({\bf k}, \omega)
\end{eqnarray}
where, $G({\bf k}, \omega) = lim_{\delta \rightarrow 0} G({\bf k}, i\omega_{n})\vert_{i\omega_{n} \rightarrow \omega + i\delta}$. $G({\bf k}, i\omega_{n})$ is the imaginary frequency transform of $\langle c_{\bf k}(\tau)c_{\bf k}(0)^{\dagger}\rangle$.  
}
\item{Order parameter, 
\begin{eqnarray}
m^{2} = \frac{1}{N}\sum_{l, i, j}\langle {\bf m}_{li}.{\bf m}_{lj}\rangle e^{i{\bf q}.({\bf r}_{i}-{\bf r}_{j})}
\end{eqnarray}
where, the index $l$ corresponds to the number of spins in the unit cell; $i$, $j$, and ${\bf r}_{ij}$ gives the 
positions of the unit cell on the triangular Bravais lattice.
}
\end{itemize}

The static path approximation has been used extensively to investigate several quantum many body phenomena, such as, 
the BCS-BEC crossover in superconductors \cite{tarat_epjb}, Fulde-Ferrell-Larkin-Ovchinnikov (FFLO) superconductivity in solid 
state systems and ultracold atomic gases \cite{mk_pra2016,mk_pra2018,mk_spinliq_prb2022,mk_prr2020}, Mott transition in frustrated lattices 
\cite{nyayabanta_prb2016,nyayabanta_jpcm2017,nyayabanta_prb2022}, d-wave superconductivity 
\cite{dagotto_prl2005}, competition and coexistence of magnetic and d-wave superconducting orders \cite{dagotto_prb2005}, 
orbital selective magnetism relevant for iron superconductors \cite{dagotto_prb2016}, strain induced superconductor-insulator transition 
in flat band lattices \cite{mk_prr2020}, heteromolecular ultracold atomic gases \cite{mk_pra2018} etc. In many of these problems the use of 
numerically exact techniques like DQMC is not feasible due either to the sign problem or to the system size restrictions (particularly for 
multiband systems). Judicial approximations are thus essential and SPA is one such approximation which can capture the low temperature 
phases and the thermal properties of these strongly correlated systems with reasonable accuracy. The technique however will 
and does fall short in situations where the physics of the ground state is almost entirely dictated by quantum fluctuations, such as, 
quantum spin liquids, heavy fermion superconductors etc. 

\begin{figure*}
\begin{center}
\includegraphics[height=12cm,width=15cm,angle=0]{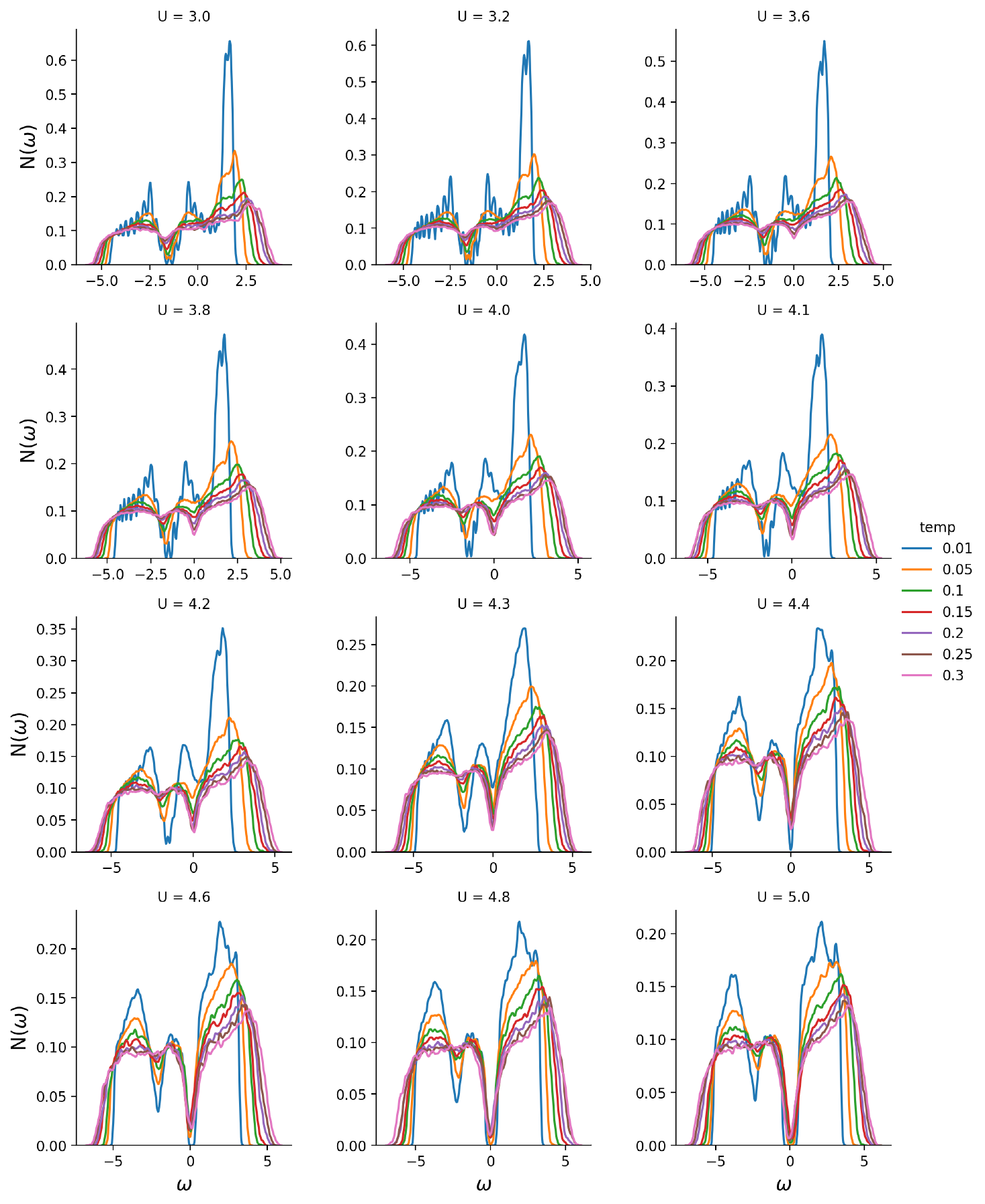}
\caption{Thermal evolution of the single particle DOS across the FI, PM, AFM and AF-MI phases. Note 
how the gapless spectra at $U < U_{c1}$ progressively develops first a soft gap followed by a Mott gap 
at the Fermi level with increasing interaction. The dominant short range magnetic correlations at the high 
temperatures are signaled by the development of a dip in the spectra at the Fermi level for $U_{c1} < U \le U_{c2}$.}
\label{fig3_suppl}
\end{center}
\end{figure*}

\textit{Precursor Coulomb phase}

Fig.\ref{fig2_suppl} shows the low temperature evolution of the precursor Coulomb phase at intermediate 
interactions in terms of the magnetic structure factor maps.  Characterized by isolated $S({\bf q})$ peaks 
giving rise to a $\sqrt{3} \times \sqrt{3}$ magnetic order at weaker interactions, the state progressively 
evolves to a Coulomb phase comprising of pinch points, with increasing interaction. Observed in spin ice 
materials such as, Ho$_{2}$Ti$_{2}$O$_{7}$, Dy$_{2}$Ti$_{2}$O$_{7}$ etc. the pinch points are salient to 
magnetic frustration and extensively investigated in the strong coupling Heisenberg limit, in the context of 
quantum magnets \cite{udagawa_prb2018,zhitomirsky_prb2008,shanon_prb2018,iqbal_prr2023,gingras_science2001,bramwell_science2009,zaharko_prb2018}. Within our choice of the numerical scheme, the system remains arrested in the Coulomb phase even at still lower 
temperatures owing to the neglect of the quantum fluctuations and the consequent absence of the thermal 
order by disorder mechanism required to stabilize a coplanar phase as the lowest energy state.  
  
The thermal evolution of the spectroscopic signatures across the interaction plane is shown in Fig.\ref{fig3_suppl}. 
The weak interaction regime is gapless at all temperatures with a finite spectral weight at the Fermi level and a robust 
high energy flat band at $\omega \sim 2t$. For $U \ge U_{c1}$ the high temperature regime develops short range 
magnetic correlations giving rise to a dip in the single particle DOS at the Fermi level, which progressively deepens 
both with the interaction and temperature as the thermal fluctuations induced short range magnetic correlations strengthen.  
The high energy flat band undergoes progressive  suppression and broadens via transfer of spectral weight.  
At $U \sim U_{c2}$ the Mott gap opens up at the Fermi level accompanied by an underlying Coulomb phase. 
The spectral signatures of the Mott insulating phase is largely immune to thermal fluctuations,  as observed in the 
strong coupling regime. 

\bibliography{kagome.bib}
    
\end{document}